\documentclass[twocolumn]{aastex631}

\shorttitle{Secular Gap Complexity in STIPs}
\shortauthors{Livesey \& Becker}

\graphicspath{{./}{figures/}}

\usepackage{physics}

\def\msun{{M_\odot}}

\newcommand\celmech{\texttt{celmech}}
\newcommand\rebound{\texttt{rebound}}
\newcommand{\revone}[1]{#1}
\newcommand{\revtwo}[1]{#1}

\begin{document}

\title{Secular Perturbations from Exterior Giants Strongly Influence Gap Complexity in Peas-in-a-Pod Exoplanetary Systems}

\correspondingauthor{Joseph R. Livesey}
\email{jrlivesey@wisc.edu}

\author[0000-0003-3888-3753]{Joseph R. Livesey}
\author[0000-0002-7733-4522]{Juliette Becker}
    \affiliation{Department of Astronomy, University of Wisconsin--Madison, Madison, WI 53706}

\begin{abstract}
It has been demonstrated that systems of tightly packed inner planets with giant exterior companions tend to have less regular orbital spacings than those without such companions. We investigate whether this observed increase in the gap complexity of the inner systems can be explained solely as the result of secular dynamics caused by the disturbing potential of the exterior companions. Amplification of mutual orbital inclinations in the inner system due to such secular dynamics may lead to the inner system attaining non-mutually transiting geometries, thereby creating artificial observed gaps that result in a higher calculated gap complexity. Using second-order secular theory, we compute time-averaged observed gap complexities along a favorable line of sight for a set of hypothetical systems, both with and without an outer giant. We find that these secular interactions can significantly contribute to the observed gap complexity dichotomy in tightly packed multiple-planet systems.
\end{abstract}

\keywords{Exoplanet detection methods (489), Exoplanet dynamics (490), Planetary system formation (1257), Planetary system evolution (2292)}

\section{Introduction}
In multiple-planet systems, planets typically have similar masses to those of their neighboring planets, and pairs of exoplanets in the same system tend to have similar orbital separations to the pairs next to them. This trend of regular spacings and similar masses was first identified by \citet{Weiss2018}, who named this class of orbital architecture the ``peas in a pod'' system. This architecture is prevalent among the present sample of transiting exoplanetary systems \citep{Millholland2017, Weiss2018, Weiss2023}, and has been proposed to be an energetically favored outcome of planet formation \citep{Adams2019, Adams2020} as well as a natural consequence of dynamical evolution thereafter \citep{Goldberg2022, Lammers2023, Ghosh2024}.

Leveraging the framework of information theory, \citet{GilbertFabrycky2020} introduced a set of quantities that parameterize the distribution of mass, orbital period, and inclination within a planetary system. These quantities are the dynamical mass, mass partitioning, monotonicity, characteristic spacing, flatness, multiplicity, and gap complexity. 
For a system of $N$ planets (and $n \equiv N - 1$ gaps), the gap complexity can be written
\begin{equation} \label{eq:gap-complexity}
    \mathcal{C} = -K_n \left [ \sum_{i=1}^n p^\star_i \log(p^\star_i) \right ] \left [ \sum_{i=1}^n \left ( p^\star_i - \frac{1}{n} \right )^2 \right ],
\end{equation}
where
\begin{equation}
    p^\star_i \equiv \frac{\log(P_{i+1}/P_i)}{\log(P_N/P_1)}.
\end{equation}
The expression in the first square brackets in Equation \ref{eq:gap-complexity} is the Shannon entropy,\footnote{\revone{Up to a factor of $\log(2)$. Here and throughout this work the natural logarithm is used.}} and the expression in the second square brackets is the disequilibrium in $p^\star_i$. $K_n$ is a normalization constant defined such that $\mathcal{C}$ has a maximum value of unity.

The gap complexity is a convex complexity, maximized where the entropy and disequilibrium are equal. Convex complexities, as measures of disorder in a physical system, are preferred over entropy or the disequilibrium alone. In particular, an ideal gas has maximum entropy despite the fact that such a system is in energy equipartition, and a perfect crystal at zero kelvins has maximum disequilibrium despite the fact that such a system has perfectly regular spatial ordering \citep{GilbertFabrycky2020}.

The gap complexity has proven a powerful metric that correlates with other physical properties of a planetary system \citep[such as planetary radii;][]{Rice2024}. Recently, \citet{HeWeiss2023} applied the gap complexity statistic to a sample of planetary systems from the Kepler Giant Planet Search \citep[KGPS;][]{KGPS} catalog, each featuring a system of tightly packed inner planets (STIP). \citet{HeWeiss2023} find a statistically significant discrepancy between the distribution of gap complexities between STIPs with and without outer giant companions (OGs). Specifically, STIPs with OGs tend to exhibit higher gap complexities.

This discrepancy potentially offers insights into the processes of planetary system formation and evolution. It suggests that accreting OGs either interrupt the development of STIPs during formation or perturb STIPs on secular timescales, increasing the system complexity. Perhaps counterintuitively, $N$-body simulations have shown that the presence of an OG during planet formation tends to decrease the gap complexity of an emerging STIP \citep{Kong2024}.

In this paper, we examine the remaining hypothesis: that the presence of an outer giant companion can secularly perturb the inner system and alter the observed gap complexity. For sufficiently large perturbations due to exterior companions \citep[e.g.,][]{Becker2017, JontoffHutter2017, Read2017}, planets in a STIP may attain non-mutually transiting configurations, leading to only a subset of a STIP's planets being seen in transit simultaneously \citep{Ballard2016, Brakensiek2016}. These dynamics will affect the gap complexity metric. We use secular theory to model STIPs with and without OGs in order to quantify their effect on the inner planets' orbital inclinations and the resulting observed gap complexity.\footnote{GitHub codebase: \href{https://github.com/jrlivesey/SecularGapComplexity}{https://github.com/jrlivesey/SecularGap Complexity}; an initial release has been deposited to Zenodo at doi:\href{https://doi.org/10.5281/zenodo.14171612}{10.5281/zenodo.14171612} \citep{SecularGapComplexity}.}
In Section \ref{sec:secdyn} we outline our model and the underlying physics. In Section \ref{sec:simulations} we detail the results of our simulations. We discuss these results in relation to previously published observational trends and address their limitations and caveats in Section \ref{sec:discussion}, and conclude in Section \ref{sec:conclude}.

\section{Secular Dynamics} \label{sec:secdyn}
\subsection{The Laplace--Lagrange Solution}
We employ the Laplace--Lagrange equations of motion, which are derived from the secular disturbing potential for an $N$-planet system. These equations describe the evolution of each orbital element in the system. We outline the procedure for calculating the inclination evolution here, following Chapter 7 of \citet{MurrayDermott1999}.

We work in an astrocentric co-ordinate system, assuming that motion due to perturbations on the star is very small, since $m_\star \gg m_j$ for every planet in the system. The first step in computing the Laplace--Lagrange inclination solution is to derive the $N \times N$ secular matrix $\vb*{B}$. This matrix has elements
\begin{align}
    B_{jj} &= -\frac{1}{4} n_j \sum_{k \neq j} \frac{m_k}{m_\star + m_j} \alpha_{jk} \bar{\alpha}_{jk} b^{(1)}_{3/2}(\alpha_{jk}) \label{eq:Bjj} \\
    B_{jk} &= \frac{1}{4} n_j \frac{m_k}{m_\star + m_j} \alpha_{jk} \bar{\alpha}_{jk} b^{(1)}_{3/2}(\alpha_{jk}), \label{eq:Bjk}
\end{align}
where $n_j$ and $m_j$ respectively denote the mean motion and mass of body $j$. For a pair of bodies $j$ and $k$, $\alpha_{jk}$ is the ratio of the outer body's semi-major axis to the inner body's semi-major axis. $\bar{\alpha}_{jk} = \alpha_{jk}$ when body $k$ is exterior to body $j$, otherwise $\bar{\alpha}_{jk} = 1$. The Laplace coefficients are defined as
\begin{equation}
    b^{(\ell)}_s(\alpha) = \frac{2}{\pi} \int_0^\pi \frac{\cos(\ell \psi) \: d\psi}{[1 + \alpha^2 - 2\alpha \cos(\psi)]^s}.
\end{equation}
We ignore the effects of stellar oblateness, but note that such effects can have a pronounced impact on the evolution of orbital inclinations in the system \citep[particularly for close-in planets as the host star evolves; e.g.,][]{Li2020, Becker2020, BrefkaBecker2021, Chen2022, Faridani2024}. The inclination frequencies $f_1, f_2, \dots, f_N$ are the eigenvalues of $\vb*{B}$, with corresponding normalized eigenvectors $\bar{\vb*{y}}_i$:
\begin{equation}
    \vb*{B} \bar{\vb*{y}}_i = f_i \bar{\vb*{y}}_i.
\end{equation}
The magnitudes of the scaled eigenvectors relevant to the problem are determined by the initial conditions of the system. We work in terms of the Poincaré co-ordinates
\begin{align}
    p_j &= I_j \sin(\Omega_j) \\
    q_j &= I_j \cos(\Omega_j)
\end{align}
and let $p_j = p_{j,0}$ and $q_j = q_{j,0}$ when we ``start the clock'' at $t = 0$. These co-ordinates are calculated for the $j$-th planet as
\begin{align}
    p_j &= \sum_{i=1}^N y_{ji} \sin(f_i t + \gamma_i) \\
    q_j &= \sum_{i=1}^N y_{ji} \cos(f_i t + \gamma_i),
\end{align}
where the $y_{ji}$ are elements of the matrix whose columns are the scaled eigenvectors, $\vb*{y}_i = T_i \bar{\vb*{y}}_i$, and the factors $T_i$ and the phase angles $\gamma_i$ are determined by the initial conditions.

At $t = 0$, we have the following system of $2N$ equations in as many unknowns:
\begin{align}
    p_{j,0} &= T_i \bar{y}_{ji} \sin(\gamma_i) \\
    q_{j,0} &= T_i \bar{y}_{ji} \cos(\gamma_i).
\end{align}
Hence we solve simultaneously for all the $T_i$ and $\gamma_i$ and obtain the linear solution for the system's secular evolution. We obtain the Laplace--Lagrange solutions using the Python package \celmech~\citep{HaddenTamayo2022}, which includes a function to quickly compute the inclination eigenvalues and scaled eigenvectors.

\subsection{Secular Response to an Outer Giant and to a Stellar Companion}
A secondary finding of \citet{HeWeiss2023} is that the presence of an outer \textit{stellar} companion (SC) does not affect the gap complexity of the STIP in the same way as an OG. We therefore also investigate the secular response of a STIP to an SC.

\begin{figure*}
    \centering
    \includegraphics[width=\linewidth]{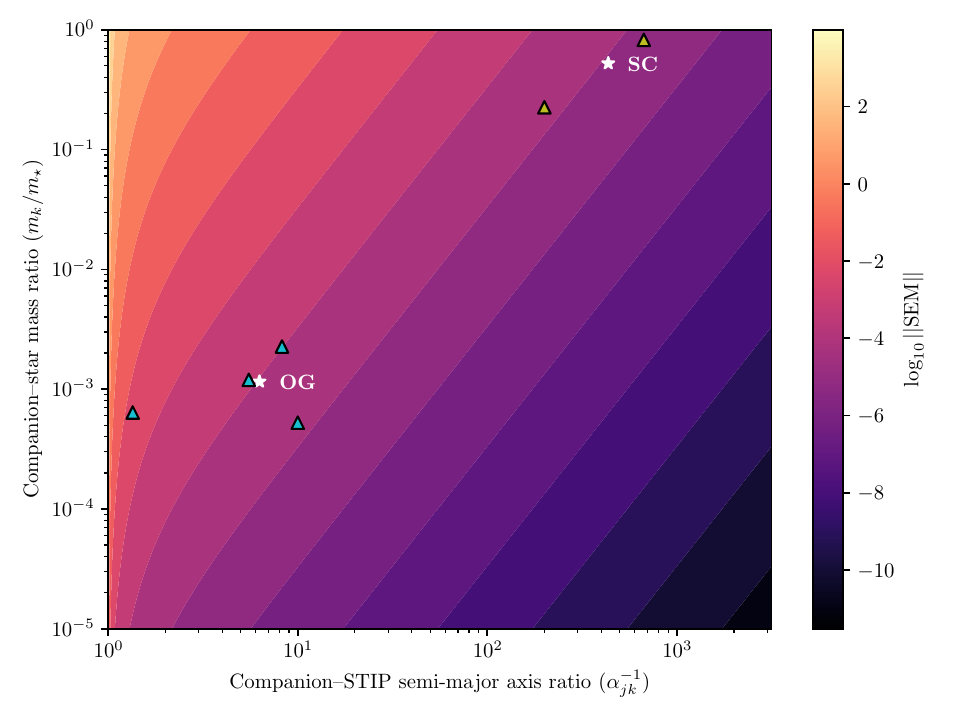}
    \caption{Factor by which we scale the secular matrix elements when moving or changing the mass of the outer companion. Overlain are KGPS systems from the \citet{HeWeiss2023} analysis containing outer companions and at least three inner planets; cyan triangles denote OG systems while olive triangles denote SC systems. \revone{The white stars indicate the averages of these two groups. On average, the scaling factor is $\sim 10^2$ times greater for OG systems than for SC systems in this sample.}}
    \label{fig:scaling}
\end{figure*}

At this point, we notice that the Laplace--Lagrange formulation appears well-suited to address this key problem, even without computing the full equations of motion for exemplar planetary systems. Note that the secular matrix elements in Equations \ref{eq:Bjj} and \ref{eq:Bjk} depend on the quantity $m_k \alpha_{jk} \bar{\alpha}_{jk} b^{(1)}_{3/2}(\alpha_{jk})$, \revone{to which we will refer as the \textit{secular effective mass} (SEM)}. Consider a 3-body system containing the primary, one small planet, and an exterior giant companion (either an OG or an SC).
In Figure \ref{fig:scaling}, we plot the locations of a fiducial OG and SC in mass--semi-major axis parameter space (given by the white points). The \revone{SEM} changes markedly when we switch from computing it using typical parameters of OGs to using typical values of SCs; it decreases by a factor of $\gtrsim 10$. The secular matrix we obtain from a SC will consequently have much smaller eigenvalues, i.e., the inclination frequencies we get are smaller.\footnote{It is worth noting that the eigenvectors themselves are not exactly preserved. In particular, $B_{12}$ and $B_{21}$ are not scaled by the same amount.}
This analysis shows why we expect smaller secular perturbations from a SC than from an OG, which may account for the observed discrepancy in inner system gap complexity between real systems containing SCs and those containing OGs.

\subsection{Setup and Parameter Space}
\label{sec:setup}
To assess the impact of an exterior companion on the dynamics of a STIP, we compute the secular evolution for idealized STIPs in the presence of outer giant companions with a range of physical and orbital parameters. We adopt a system of units in which $G = 1$ and the mass of the primary is $m_\star = 1$. Then, Kepler's third law becomes $P = 2\pi a^{3/2}$, with both the orbital period $P$ and semi-major axis $a$ being dimensionless quantities.

Each model planetary system in our sample contains $N + 2$ bodies: (i) the primary, (ii) a STIP comprising $N$ planets, and (iii) an outer companion.
For each drawn set of initial conditions, two geometries are modeled: one with the outer companion, and a second control geometry with no outer companion. 

\begin{figure}
    \centering
    \includegraphics[width=\linewidth]{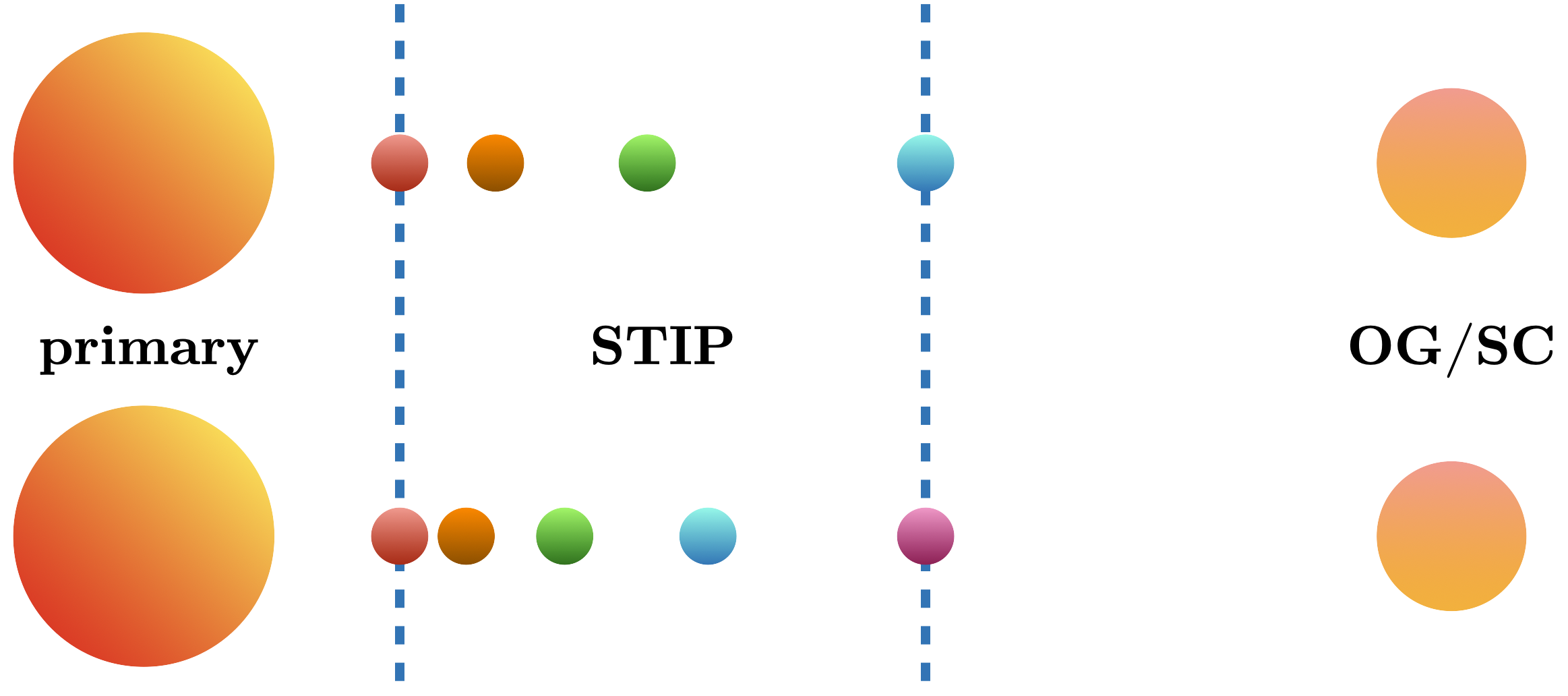}
    \caption{Illustration of the dependence of our systems' architecture on the STIP multiplicity. \revone{The vertical lines indicate the fixed values of $a_1$ and $a_N$.} Distances are not to scale. \textit{Top:} The $N = 4$ case. \textit{Bottom:} The $N = 5$ case.}
    \label{fig:stip-setup}
\end{figure}

\revone{\citet{Weiss2018} find that planets in a ``peas-in-a-pod'' system obey the spacing relationship
\begin{equation} \label{eq:ratioratio}
    \mathcal{P} \equiv \frac{P_{j+1}/P_j}{P_j/P_{j-1}} = 1.03 \pm 0.27.
\end{equation}
In our simulations, we assign semi-major axes to the STIP planets according to this relationship. Note that $\mathcal{P}^{2/3} = (a_{j+2}/a_{j+1})/(a_{j+1}/a_j)$ by Kepler's third law. We fix $a_1$ and $a_N$, then find the semi-major axis of each intermediate planet by using this spacing parameter. Re-casting Equation \ref{eq:ratioratio} as
\begin{equation}
    \frac{a_{j+1} a_{j-1}}{\mathcal{P}^{2/3} a_j^2} = 1,
\end{equation}
we have $N - 2$ equations, in as many unknowns, for the remaining semi-major axes.
Going forward, we assume that $\mathcal{P} = 1$, which is nearly the mean observed value of \citet{Weiss2018} and allows us to re-write this spacing relationship as
\begin{equation} \label{eq:linearized}
    \ell_j = \tfrac{1}{2} (\ell_{j-1} + \ell_{j+1}),
\end{equation}
where $\ell_i \equiv \log(a_i)$. We realize Equation \ref{eq:linearized} as a matrix equation, whence we notice that the vector of logarithmic distances is an eigenvector --- with unit eigenvalue --- of the $N \times N$ matrix $\vb*{\Lambda}$ that has elements
\begin{align}
    \Lambda_{jk} &= \delta_{j,1} \delta_{k,1} + \delta_{j,N} \delta_{k,N} \nonumber \\
    &\quad
    + \tfrac{1}{2} (1 - \delta_{j,N}) \delta_{j,k+1} + \tfrac{1}{2} (1 - \delta_{j,1}) \delta_{j,k-1}
\end{align}
where $\delta_{j,k}$ is the Kronecker delta.
We re-scale these solutions in order to match our selected innermost and outermost semi-major axes.
We choose $a_1 = 0.1$ and $a_N = 0.5$.}
The exact distances are somewhat arbitrary; the important aspect of this setup is that our idealized STIP's true gap complexity is zero. A cartoon illustration of this setup is given in Figure \ref{fig:stip-setup}.

Each STIP planet has a mass of $10^{-5}$: characteristic of an Earth orbiting an M dwarf or a super-Earth orbiting a G dwarf. 
The initial inclinations are sampled from the distribution fit empirically by \citet{Fabrycky2015}:
\begin{equation} \label{eq:stip-incs}
    I_{j,0} \sim \text{Rayleigh}(2.5^\circ).
\end{equation}
We assume zero eccentricity for each body and set them to be initially nodally aligned, $\Omega_{j,0} = 0$.

The Laplace--Lagrange equations constitute a fully linear model, and our approach is thus computationally inexpensive. Enabled by the ease of solving the planetary equations of motion, we probe a parameter space with several dimensions:
\begin{itemize}
    \item The mass of the outer companion, which we vary uniformly between $10^{-3}$ and \revone{0.5}.
    \item The semi-major axis of the outer companion, which we vary uniformly between \revone{0.75 and 10}.
    \item The multiplicity $N$ of the STIP.
\end{itemize}

The gap complexity of a planetary system is evaluated based upon the orbital periods of all \textit{observed} bodies, and depends strongly on their spacings. Using the secular approach previously described, we set up STIP-like planetary systems and take account of which planets are observable in transits at any given time along a chosen line of sight.
It is very straightforward to set a limit on the range of sky-plane inclinations in which a planet is transiting. Since we choose planetary orbits with $e = 0$, a planet at orbital distance $a$ is always non-transiting (along a line of sight perpendicular to the line of nodes) if its inclination satisfies
\begin{equation}
    I > \arctan(R_\star / a),
\end{equation}
where $R_\star$ is the stellar radius.

In our simulations, the line of sight along which we evaluate the observed gap complexity is always orthogonal to the line of nodes. The line of sight tracks the median inclination plane; i.e., the plane normal to the total orbital angular momentum of all planets in the STIP. Thus, the gap complexity time series we obtain, $\tilde{\mathcal{C}}(t)$, is not the gap complexity that would be obtained by a single observer (e.g., an astronomer on Earth). It is instead the value obtained along a line of sight orthogonal to the line of nodes and in the mean inclination plane of the STIP.\footnote{This assumption allows us to ignore any overall precession/recession of the STIP plane about the line of nodes. Effectively, it eliminates a mode of inclination variability among the STIP planets (sets one of the $f_i = 0$).} 

We choose to compute the gap complexity from this geometry to better compare with observational results. The gap complexity metric does not apply to systems of two or fewer planets, and so analyses in the literature that relate planet properties to gap complexity are sampling only the systems where multiple planets are seen to transit. While a multi-planet system could be seen as a single-planet system due to geometric effects \citep{Ballard2016}, those systems would not be identified as multi-planet systems for which gap complexities can be computed. As a result, we restrict our analysis to the lines of sight from which our idealized STIPs are most likely to be identified as $N > 2$-planet systems.
Ignoring secular changes in the longitudes of ascending node, this line of sight is that along which we will observe the lowest value of the gap complexity on average, since it is where most planets will be found to be transiting. It is also the line along which we will observe the most time-dependent behavior of the gap complexity, as planets come in and out of the transiting plane but the gap complexity will be defined (which requires at least three planets to be seen in transit) with the highest probability. 

\revone{
\subsection{Validity of the Secular Approximation}
Under second-order secular theory, the disturbing potential experienced by a planet in our system does not include any mixed terms that would couple eccentricities with inclinations. This framework is therefore only of use in describing systems in which planets are sufficiently distant from one another, are not in any mean motion resonances, and do not exchange energy between their eccentricity and inclination oscillation modes \citep{MurrayDermott1999}. There is no general analytic criterion that determines whether or not a system's dynamics are well described by second-order secular theory; its validity must be tested by comparison of results with $N$-body simulations.}

\revone{To that end, we ran multiple $N$-body simulations of one of our simulated systems, in the regime where either the OG is very massive or orbits close to the outermost planet in the STIP.\footnote{These results can be found on this paper's GitHub repository.} These simulations were run with the \rebound~$N$-body simulation suite\footnote{\rebound~is open-source and is available at \href{https://github.com/hannorein/rebound}{https://github.com/ hannorein/rebound}. Our simulations are performed with v4.3.0.} \citep{Rein2012} using \texttt{whfast}, a symplectic Wisdom--Holman integrator \citep{Wisdom1991, Rein2015}. We find good qualitative agreement between the secular approximation and the $N$-body results up to the Hill stability boundary: $a_\text{OG} - a_\text{STIP} = 2\sqrt{3} R_H$, where $R_H$ is the planets' mutual Hill radius \citep{Gladman1993}.
}

\section{Results of Simulations} \label{sec:simulations}
For each simulation, we take the time-averaged gap complexity $\langle \tilde{\mathcal{C}} \rangle$ as our metric to assess the influence of the OG.
Observations of systems in the \textit{Kepler} sample represent only snapshots in their histories; we have only instantaneous measurements of their gap complexities.
With larger samples of multi-planet systems, however, the distribution of gap complexities will approximate the time-averaged values. Consequently, if secular evolution due to an outer giant is a viable explanation for the observed statistical differences in measured gap complexities between systems with and without outer giants, then the observational correlation should align with differences in systems' modeled, time-averaged gap complexities. 

To determine these differences, we compute the secular evolution for \revone{2,500} sets of initial conditions following the procedure described in Section \ref{sec:setup}. First, we set the STIP up to have $N = 4$ planets with zero initial gap complexity and draw orbital inclinations for each planet in the STIP. Then, we create two realizations of this STIP: one with an OG with parameters drawn from a grid with the extent described in Section \ref{sec:setup} and an orbital inclination of $I_\text{OG} = 10^\circ$ relative to the plane of the STIP, and one without an OG. We then construct similar ensembles for different STIP multiplicities and relative inclinations. For each set of initial conditions, we compute the inclination evolution of the planets in the STIP and the gap complexity.
The results are shown in Figure \ref{fig:heatmaps}. The horizontal axes in these heatmaps are the ratio in semi-major axis between the OG and the outermost STIP planet (the reciprocal of the parameter $\alpha$). We find that changing the relative inclination has a much stronger effect than changing the multiplicity of the STIP on the gap complexity. 

\begin{figure*}
    \centering
    \includegraphics[width=\linewidth]{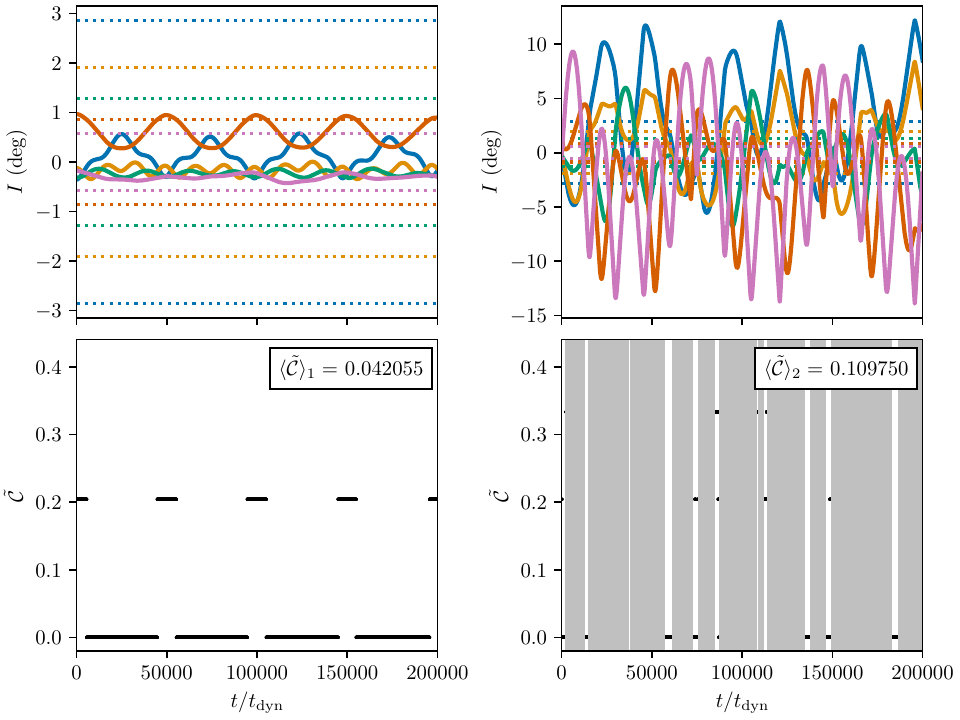}
    \caption{A pair of simulations --- one with an exterior companion and one without --- of a STIP containing five planets. The companion in this case has a mass of $10^{-2}$, lies \revone{at $a = 2$,} and has an inclination of $10^\circ$ relative to the STIP plane. The Laplace--Lagrange solution is computed out to \revone{$10^5$ times} the shortest dynamical timescale in the system, $t_\text{dyn}$ (effectively the orbital period of the innermost planet). \revone{Note that this plot does not capture all secular cycles; the short time span here is chosen for ease of interpretation.} For this particular system, the presence of the OG \revone{increases} $\mathcal{C}$ on average. \textit{Left:} The inclinations of all five STIP planets over time, with the maximum transiting inclinations for each planet indicated with the dotted lines of corresponding colors. \textit{Right:} The evolution of the gap complexity along our chosen line of sight. Vertical grey bars indicate spans of time at which the gap complexity is undefined (i.e., there are only 0, 1, or 2 planets in the transiting plane).}
    \label{fig:simpair-1}
\end{figure*}
\begin{figure*}
    \centering
    \includegraphics[width=\linewidth]{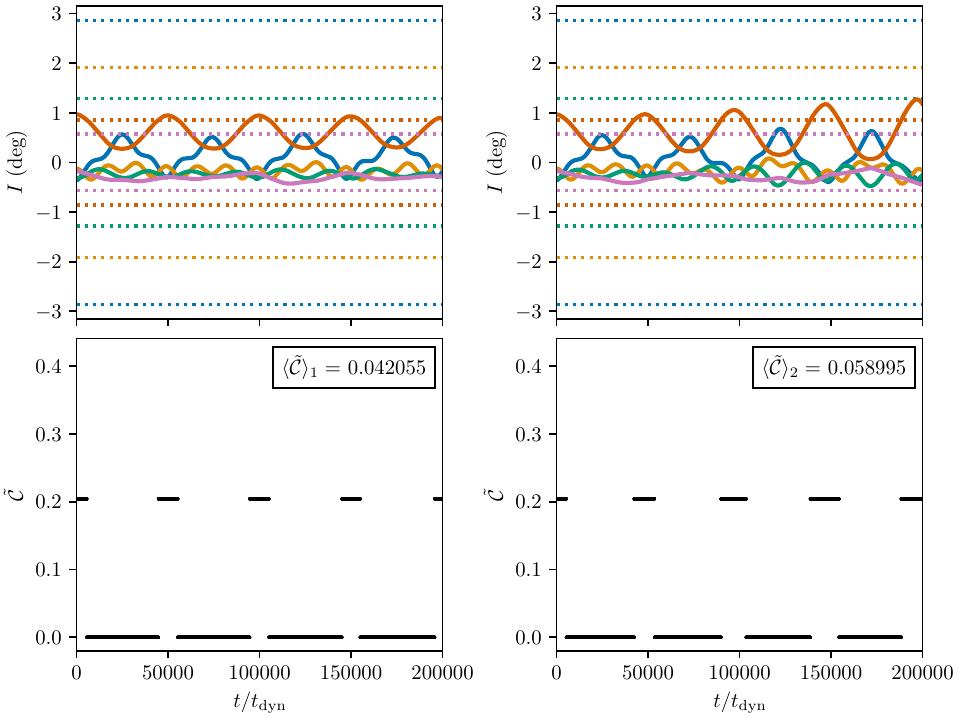}
    \caption{Same as Figure \ref{fig:simpair-1}, but with a companion of mass $10^{-3}$ that lies at $a = 10$. In this case, the presence of the companion results in a small increase in the average gap complexity. The fact that the companion mass is small and its semi-major axis is large places it in the low-$m/m_\star$ and high-$\alpha^{-1}$ region of the parameter space shown in Figure \ref{fig:scaling}, where we expect little enhancement in the secular forcing.}
    \label{fig:simpair-2}
\end{figure*}

We show the two extremes of a STIP's response to the presence of an OG in Figure \ref{fig:simpair-1} and Figure \ref{fig:simpair-2}.
In Figure \ref{fig:simpair-1}, we show a case with a massive ($10^3$ times a STIP planet's mass), close-in ($\alpha \simeq 1/2)$ OG.\footnote{From here on, $\alpha \equiv a_\text{out} / a_\text{OG}$, where $a_\text{out}$ is the orbital distance of the outermost planet in the STIP and $a_\text{OG}$ that of the outer giant.} In the top panels of Figure \ref{fig:simpair-1}, we show (left panel) the inclination evolution of the STIP and (right panel) the computed gap complexity, where greyed out regions correspond to systems where gap complexity cannot be computed because less than two planets transit. \revone{The time-averaged value $\langle \tilde{\mathcal{C}} \rangle$ is calculated only over the ranges in time during which the gap complexity is mathematically defined.} For this massive, close-in outer giant, the system is usually not seen as a high-multiplicity ($N \geq 3$) system, and when at least three planets transit the computed gap complexity has a time-averaged value of 0.11. In contrast, the bottom panel shows the computed secular evolution of the STIP with the same initial conditions, but without the outer giant. In both of these figures and hereafter, the subscript ``1'' on the time-averaged $\tilde{\mathcal{C}}$ indicates a system without an OG, while the subscript ``2'' indicates its counterpart with an OG. Without the OG, STIP planets are brought out of the transiting plane infrequently \citep[as generally expected in a system of roughly coplanar planets;][]{Becker2016}, leaving the observed gap complexity zero most of the time.  In this case, the time-averaged gap complexity is 0.03, lower than in the case with identical initial conditions but including the outer giant. The additional eigenmode introduced by the OG entirely alters the dynamical character of the system, frequently bringing multiple planets out of the transiting plane and changing the observed gap complexity.

An alternative scenario is shown in Figure \ref{fig:simpair-2}. Here, the companion is smaller (only $10^2$ times the STIP planets' masses) and farther out, at $\alpha \simeq 1/10$. The OG's impact on the inner system's gap complexity in this case is negligible, with the inclination evolution between the two cases with and without the OG being almost identical with only minor changes in amplitude.

In addition to the \revone{2,500} iterations with $N = 4$ and $I_\text{OG} = 10^\circ$ described above, we also run four additional sets of simulations (with \revone{2,500} sets of initial conditions each, for a total of \revone{12,500} unique simulations) to assess the impact of the number of STIP planets and the inclination of the OG on the gap complexity variation. These four additional simulation sets vary the number of planets in the STIP ($N = 4, 5, 6$) and \revtwo{the initial relative inclination of the OG ($I_\text{OG} = 10^\circ, 20^\circ, 30^\circ$).} The results of these ensembles are shown in Figure \ref{fig:heatmaps}. We see that the greatest amplification in the gap complexity occurs when the outer companion is close-in and large. By comparing the ensembles, we see that \revtwo{the number of planets in the STIP influences the gap complexity amplification more than the relative inclination between the OG and the STIP.} The structure seen in each of these heatmaps indicates that the magnitude and sign of the gap complexity change induced by the addition of an outer companion depends strongly on its mass and orbital parameters.
\def\figsize{0.47}
\revone{
\begin{figure*}
    \gridline{
        \fig{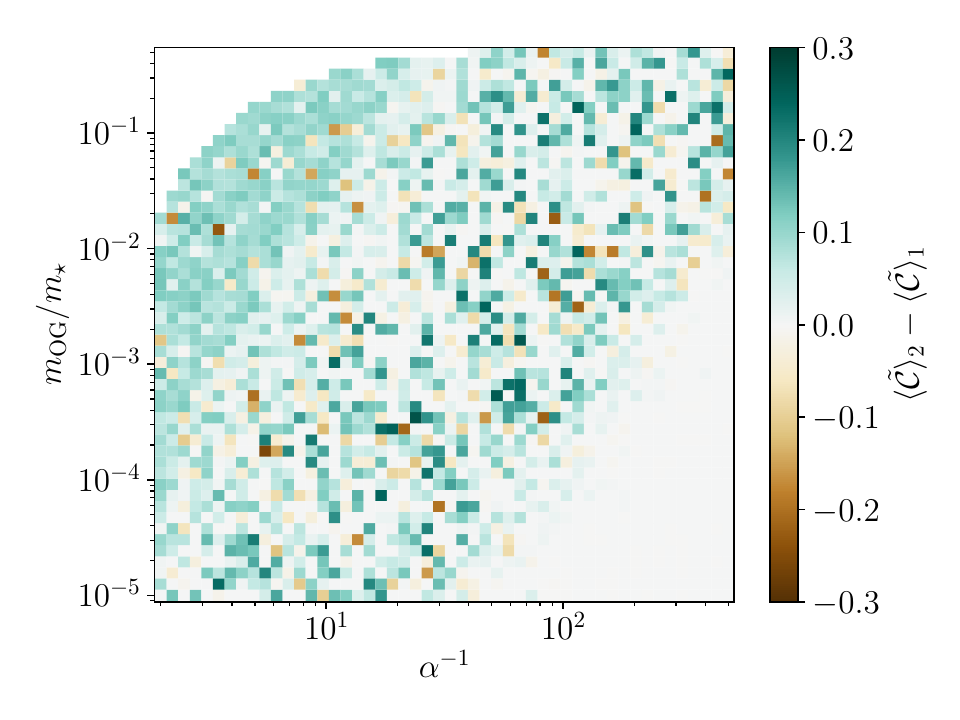}{\figsize\textwidth}{(a) $N = 4$, $I_\text{OG} = 10^\circ$. \revone{Average change in $\langle \tilde{\mathcal{C}} \rangle$: $+0.036$.}}
    }
    \gridline{
        \fig{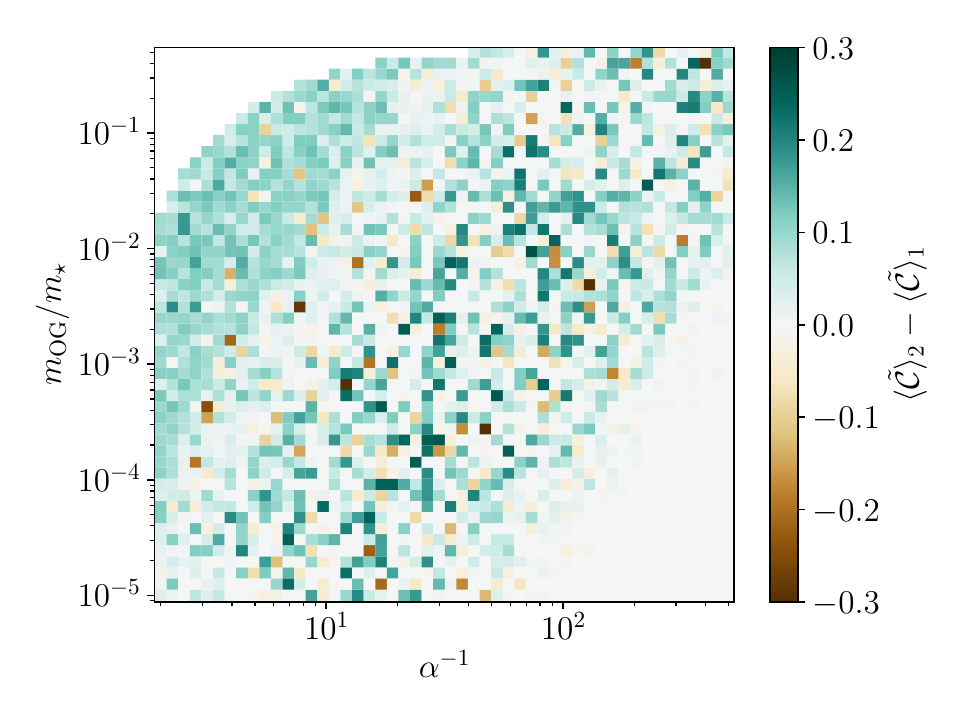}{\figsize\textwidth}{(b) $N = 4$, $I_\text{OG} = 20^\circ$. \revone{Average change in $\langle \tilde{\mathcal{C}} \rangle$: $+0.04$.}}
        \fig{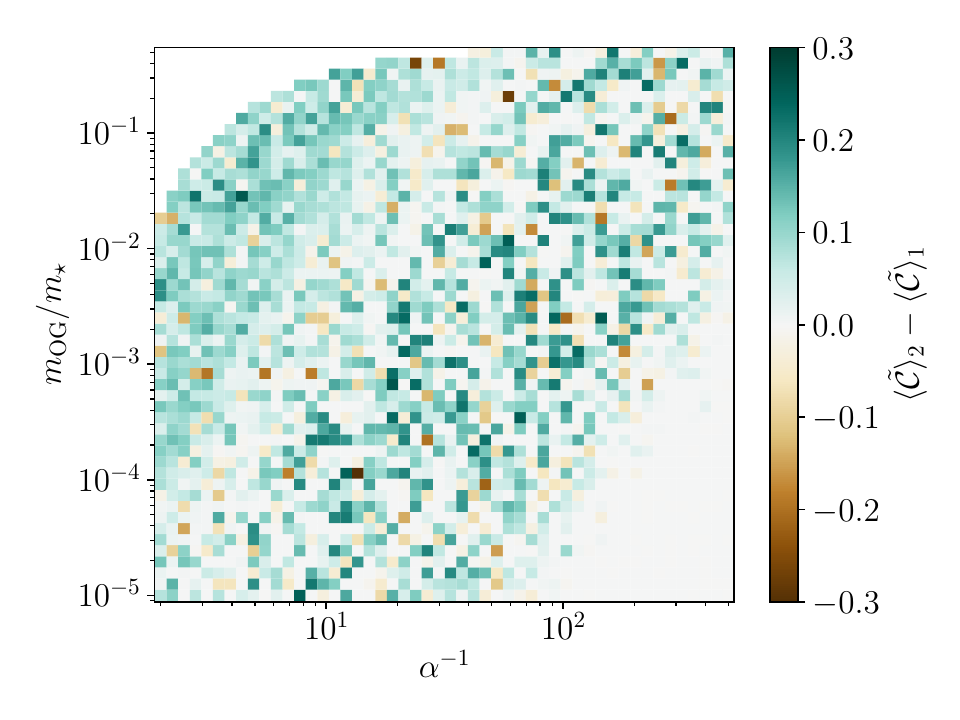}{\figsize\textwidth}{(c) $N = 4$, $I_\text{OG} = 30^\circ$. \revone{Average change in $\langle \tilde{\mathcal{C}} \rangle$: $+0.042$.}}
    }
    \gridline{
        \fig{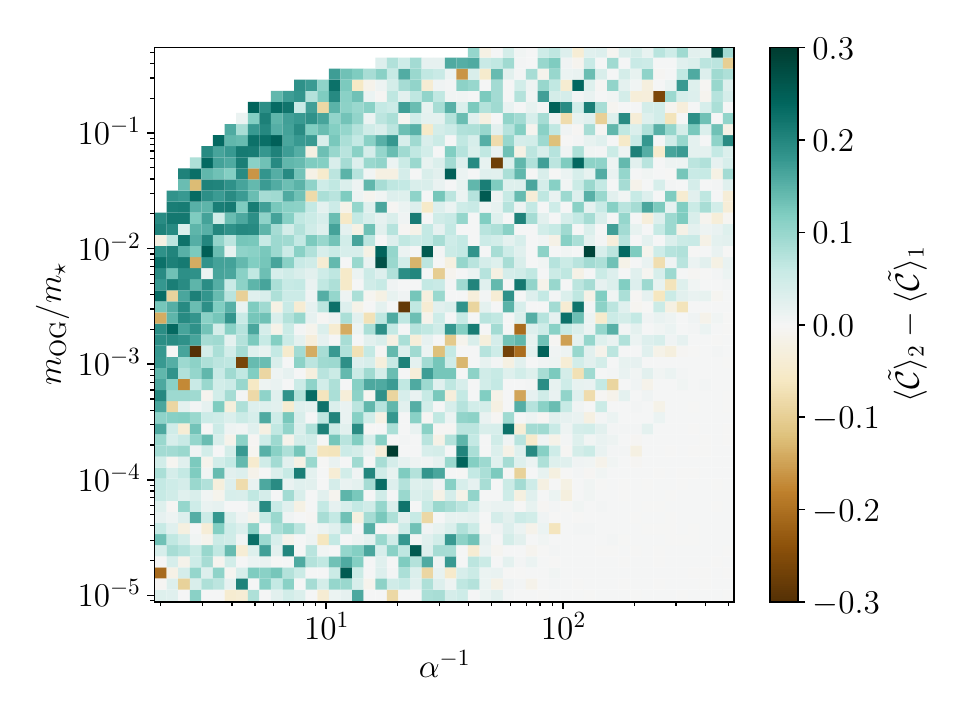}{\figsize\textwidth}{(e) $N = 5$, $I_\text{OG} = 10^\circ$. \revone{Average change in $\langle \tilde{\mathcal{C}} \rangle$: $+0.05$.}}
        \fig{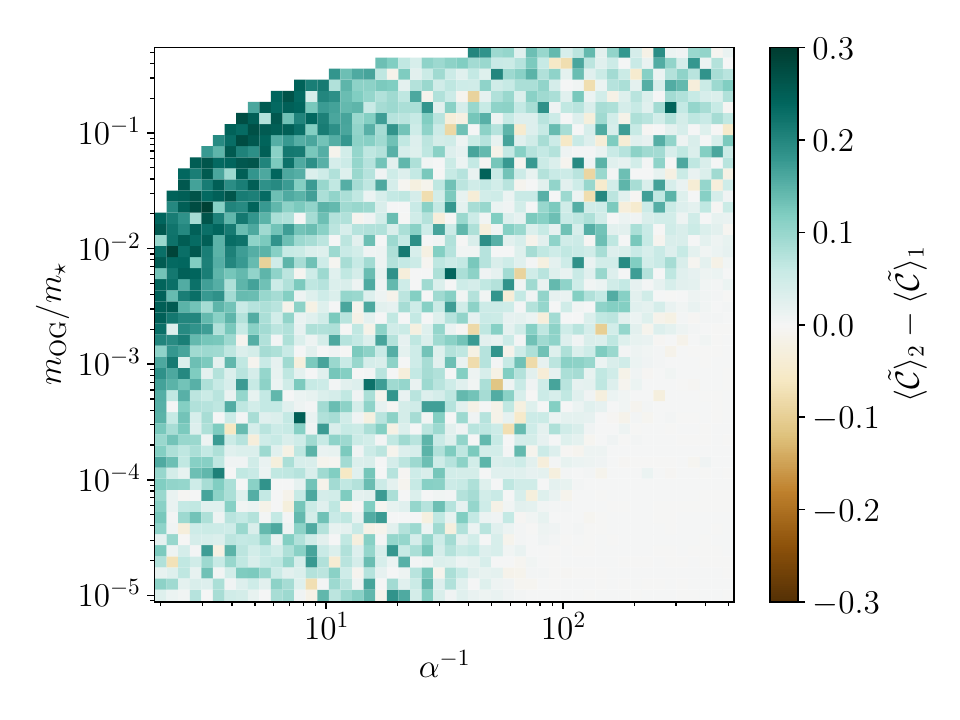}{\figsize\textwidth}{(f) $N = 6$, $I_\text{OG} = 10^\circ$. \revone{Average change in $\langle \tilde{\mathcal{C}} \rangle$: $+0.062$.}}
    }
    \caption{Change in $\langle \tilde{\mathcal{C}} \rangle$ for a sample of possible system architectures. At each point in these parameter spaces lies a pair of simulations, one of which contains a companion and the other of which does not, that are otherwise identical. \revtwo{Each simulation is run for $10^9$ times the orbital period of the innermost planet.} \revone{The blank regions at low orbital separations and high companion masses denote Hill unstable configurations.} The introduction of an exterior giant companion to the system affects the gap complexity in a way that is dependent not only on its mass and orbital distance, but also on its relative inclination to the STIP plane and to the number of planets in the STIP.}
    \label{fig:heatmaps}
\end{figure*}
}

\section{Discussion} \label{sec:discussion}
An analysis performed by \citet{HeWeiss2023} identified a new, statistically significant dichotomy in the \textit{Kepler} + KGPS sample: planetary systems containing a gas giant exterior to a system of smaller planets possess more irregular orbital spacings. This dichotomy could potentially be explained either by dynamics driven by the companions during planet formation \citep[e.g.,][]{Kong2024} or by post-formation dynamics. 

Using second-order secular theory, we have investigated the effect of outer gas giant companions on the observed orbital spacings within a system of tightly packed inner planets (STIP). Specifically, we looked at the evolution of each planet's inclination angle with and without an outer giant (OG), and assessed the observed gap complexity of the system in both scenarios as a function of time. The gap complexity is assessed along a particular, time-evolving line of sight: the line perpendicular to the line of nodes and within the total angular momentum plane of the STIP. This line of sight is chosen because it is the most favorable angle from which the highest number of planets would be discovered. 
We find that:
\begin{itemize}
    \item Secular interactions between the OG and the STIP planets in such a system lead to amplifications of planets' inclinations, bringing them out of the transiting plane more of the time and overall increasing the average observational gap complexity.
    \item These secular oscillations could account for the statistical trends derived by \citet{HeWeiss2023}: that OGs amplify gap complexity while SCs do not.
    \item The degree to which and the direction in which the gap complexity is modulated by the presence of an outer companion is dictated primarily by the physical (mass) and orbital parameters (semi-major axis and orbital inclination) of the companion.
\end{itemize}

Our results suggest that the observed dependence of gap complexity on the presence of exterior giant companions in the \textit{Kepler} sample can be  explained using secular theory. However, we emphasize that this finding does not preclude additional explanations for this trend.

\subsection{Comparison with Observational Trends: SCs vs. OGs}
\citet{HeWeiss2023} perform a Kolmogorov--Smirnov (K--S) test on their data and find that systems containing OGs have a different distribution in $\mathcal{C}$ from those without, with $p = 0.017$. Conversely, they find that the set of systems containing \textit{either} an OG or SC do not have a different distribution in $\mathcal{C}$ from those without, with a K--S $p = 0.25$. They conclude that OGs affect an inner system's gap complexity, while SCs do not.

We have established that accounting for secular oscillations, for certain combinations of OG parameters (at high mass and low $\alpha^{-1}$), the gap complexity of the inner system increases with the introduction of the OG. The regions of parameter space in which the OG depresses the gap complexity are much smaller. As such, secular interactions between a STIP and an OG tend to increase the STIP's gap complexity. Provided, then, that the planetary systems in the \citet{HeWeiss2023} sample fall within regions of parameter space in which we expect large enhancement in the secular matrix elements, our secular oscillations explanation is consistent with their derived gap complexity trend. In Figure \ref{fig:scaling} we overlay the points in OG mass--semi-major axis space occupied by the companion planets in the KGPS sample used by \citet{HeWeiss2023}, and provide these observed parameters for the reader's convenience in Table \ref{tab:he-weiss-sample}. The ``STIP + OG'' systems from their sample do fall within these favorable regions of parameter space, while the ``STIP + SC'' systems do not. Therefore, the gap complexity amplification through secular oscillations is consistent not only with the increased observational gap complexity in systems containing outer giant companions, but also the lack of such an effect under the influence of a stellar companion.

\begin{deluxetable}{cc|ccc|c}
\tablenum{1}
\tablecaption{KGPS systems containing an inner system of $\geq 3$ planets and an outer giant companion \citep{KGPS}.} \label{tab:he-weiss-sample}
\tablewidth{0pt}
\tablehead{
\colhead{KOI} & \colhead{\textit{Kepler} No.} & \colhead{$m_\star$} & \colhead{$m_\text{OG} \sin(i)$} & \colhead{$\alpha$} & \colhead{Type} \\
\colhead{} & \colhead{} & \colhead{($\msun$)} & \colhead{($M_J$)} & \colhead{} & \colhead{}
}
\startdata
85 & 65 & 1.24 & 0.673 & 0.100 & OG \\
148 & 48 & 0.91 & 2.162 & 0.121 & OG \\
316 & 139 & 1.08 & 1.353 & 0.181 & OG \\
351 & 90 & 1.11 & 0.743 & 0.742 & OG \\
2169 & 1130 & 0.94 & 221.7 & 0.005 & SC \\
3158 & 444 & 0.73 & 630 & 0.001 & SC
\enddata
\tablecomments{Here $i$ denotes the sky-plane orbital inclination. We adopt the minimum masses $m_\text{OG} \sin(i)$ for the outer companions as true masses in Figure \ref{fig:scaling}. We thus obtain mass estimates that are accurate to within an order of magnitude. $\alpha$ is the ratio between the semi-major axes of the outermost STIP planet and the exterior companion.}
\end{deluxetable}

\subsection{Limitations of Gap Complexity as an Observational Metric}
In the exoplanet census, there appears to be a correlation between the presence of inner super-Earths and outer giant companions \citep{Zhu2018, Bryan2019, Bryan2024}. This correlation indicates that the dynamics we describe in this work may operate often. 
In extreme cases (such as the example shown in the top panel of Figure \ref{fig:simpair-1}), this may lead to a STIP being seen as a one- or two-planet system, in which case the system's underlying architecture (though truly consistent with the ``peas in a pod'' paradigm) will be incorrectly classified. 
More often, one or two planets may be non-transiting, resulting in higher measured gap complexity. In rare cases, these non-transiting planets may be later discovered via radial velocity observations \citep[e.g.,][]{Buchhave2016}.
Such systems may have gap complexities that are presently incorrectly calculated, resulting in biases in analyses that use gap complexity as a metric. 
Another potential limitation in accurately determining gap complexities is the presence of planets that are too small to be detected. Systems with these undetected planets tend to exhibit higher average gap complexities. \citet{Thomas2024} provides a detailed analysis of how the absence of these small planets can influence the observed gap complexity in an example system. These factors must be considered when dynamical conclusions are drawn from observationally determined gap complexities.

\subsection{Directions for Future Work}
We have addressed scenarios in which the presence of an OG amplifies the gap complexity of an inner system. \revtwo{While the general trend seen in Figure \ref{fig:heatmaps} is that more massive, close-in outer giants increase gap complexity, there are individual simulations where the presence of an OG decreases the gap complexity. This regime is not particularly common, but its presence is noteworthy, as it indicates that while population-wide values of gap complexity are meaningful, individual systems may have physical effects with the opposite trend seen in the population as a whole.}

\revone{As we have discussed, our treatment of the gap complexity problem using second-order secular theory ignores any coupling between the orbital eccentricities and inclinations in a system. In particular, higher-order effects like Lidov--Kozai oscillations due to an OG will also act to modify the average gap complexity of a STIP on sufficiently long timescales.}

\section{Conclusions}
\label{sec:conclude}
Our secular theory analysis reveals an insight into the role of outer gas giant companions in modulating the orbital spacings within systems of tightly packed inner planets, the class of systems from which the ``peas in a pod'' archetype was derived. We find that the presence of an outer giant planet can generally lead to increased inclination oscillation amplitudes of the inner planets, causing them to attain configurations in which some planets are non-transiting more frequently. 
This interaction results in an overall increase in the observational gap complexity, aligning with the statistical trends identified by \citet{HeWeiss2023}. The degree of this modulation is largely influenced by the mass, semi-major axis, and orbital inclination of the outer giant planet.
This theoretical framework can explain the observed dichotomy in the \textit{Kepler} + KGPS sample, where planetary systems with exterior gas giants exhibit more irregular orbital spacings compared to those with more distant stellar companions or with no companions at all. \\

\section*{Acknowledgments}
J.R.L. acknowledges financial support from a Fluno Graduate Fellowship through the University of Wisconsin--Madison.
\revone{The authors also acknowledge the Wisconsin Center for Origins Research (WiCOR) and the Office of the Vice Chancellor for Research and Graduate Education (OVCRGE) at the University of Wisconsin--Madison for their support. This research was made possible in part by funding provided through the OVCRGE's Research Forward program. }

We thank Matthias He and Lauren Weiss for useful conversations\revone{, and further thank the anonymous reviewer whose suggested revisions greatly improved this manuscript.}

\software{Astropy \citep{astropy2022}, Celmech \citep{HaddenTamayo2022}, Matplotlib \citep{Hunter2007}, Numpy \citep{Harris2020}, Rebound \citep{Rein2012}, Seaborn \citep{Waskom2021}.}

\bibliography{references}{}
\bibliographystyle{aasjournal}

\end{document}